\documentclass[11pt]{article}
\usepackage{moriond,epsfig,amsfonts,amssymb} 

\newcommand{\mrm}[1]{\mbox{\rm #1}}

\newcommand{\be}{\begin{equation}}
\newcommand{\bea}{\begin{eqnarray}}
\newcommand{\eea}{\end{eqnarray}}
\newcommand{\ee}{\end{equation}}
\newcommand{\rfn}[1]{(\ref{#1})}
\newcommand{\eq}[1]{Eq.~(\ref{#1})}
\newcommand{\D}{D\hspace{-8pt}\slash}
\newcommand{\dv}{\partial\hspace{-7pt}\slash}

\newcommand{\dvr}{\stackrel{\hspace{3pt}\rightarrow}{\partial\hspace{-7pt}\slash}}

\newcommand{\gev}{ {\rm GeV} }

\begin{document}
\vspace*{4cm}
\title{An effective theory for Leptogenesis}

\author{A. Broncano}

\address{ Dept. de F\'{\i}sica Te\'orica, C-XI, Facultad de Ciencias, 
Univ.~Aut\'onoma de Madrid, Cantoblanco, 28049 Madrid, Spain}

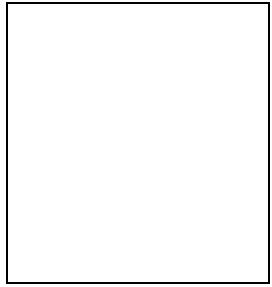
\maketitle
\abstracts{The effective Lagrangian for the seesaw model is derived. 
Besides the usual dimension-5 operator responsible for light neutrino masses, a dimension-6 operator is included which, for three or less heavy neutrino generations, provides a generic link between low-energy observables and all physical parameters of the high-energy seesaw Lagrangian. Leptogenesis can be then confirmed or ruled out through the measurement of 
neutrino masses and mixings and the exotic neutrino couplings originated from the dimension-6 operator.}

\section{Introduction}

Here, we present the work of Ref~\cite{BGJ} in which we derived the effective Lagrangian of the seesaw model. The seesaw model~\cite{seesaw} itself provides a natural explanation for both the puzzle of neutrino masses and the baryon asymmetry of the Universe, two experimental evidences for which the Standard Model (SM) cannot account. 
The model is just based on introducing in the SM Lagrangian gauge singlet fermions with $L$-violating Majorana masses of order ${\cal M} \gg M_W$. The singlet fermions couple to the massless weak doublet neutrinos of the SM and to the Higgs boson through Yukawa interactions.
Upon spontaneous symmetry breakdown of the electroweak gauge symmetry, the otherwise massless weakly-interacting neutrinos develop small masses 
$\sim -m_{\rm Dirac}^2/{\cal M}$. Present neutrino oscillation
data~\cite{experiments} indicates a seesaw scale ${\cal M}$ of new physics of ${\cal O}(10^{16} \gev)$ and at least two heavy Majorana singlet neutrinos. As soon as two or more Majorana  neutrinos are present in the seesaw model, an attractive scenario opens up  for solving the puzzle of the matter-antimatter asymmetry of the universe: leptogenesis~\cite{fy} generated  by decay of the
heavy Majorana neutrinos into light 
leptons and the Higgs boson at  the scale ${\cal M}$. The violation  of $L$, $CP$ and $C$ symmetries in the decays and the out-of-thermal-equilibrium situation produced by the expansion of the universe, naturally  provide all the conditions to generate an excess of lepton density~\cite{Sakharov}. The SM interactions recycle about half of this lepton asymmetry into a baryon asymmetry by 
active sphaleron processes~\cite{sphaleron}.  The seesaw model is, therefore, an extremely elegant and highly economical explanation for light neutrino masses and the cosmological baryon asymmetry. 

Establishing whether light neutrino masses are the result of the seesaw model requires finding an experimental signature beyond the existence of light neutrino masses. The discovery of the Majorana nature of the neutrino field by the observation, for instance, of neutrinoless double beta decay~\cite{0nu2beta} would be a major breakthrough. That, added to a possible experimental measurement of $CP$-violation in the lepton sector, would confirm leptogenesis in a promising candidate for explaining the matter-antimatter asymmetry of the Universe. It is then of prime importance to determine what is the connection, if any, between the parameters of the high-energy seesaw Lagrangian and those to be measured in on-going or future experiments. 

We present the effective Lagrangian of the seesaw theory with the minimal set of higher-dimensional operators which are necessary to establish a generic relationship between low-energy observables and the leptogenesis parameters~\cite{BGJ}.

\section{The seesaw effective Lagrangian } 

We consider the minimal extension of the Standard Model with $n$ light generations in which
$n'$ right-handed neutrinos $N_R$ are added to the field content.  

The most general gauge invariant renormalizable Lagrangian is given by 
\be
\label{Lagrangian}
{\cal L} = {\cal L}_{SM} + i \overline{N_R}\, \dv \ N_R
-  \overline{\ell_L} \,{\widetilde\phi} \, {Y_\nu}  \, N_R  
-\frac{1}{2} \overline{{N_R}^c} \, M \,{N_R} 
+ \mrm{h.c} ~.
\ee
where $ {\cal L}_{SM}$ is the SM Lagrangian. 

Since the right-handed neutrinos $N_R$ are $SU(3)_C \times SU(2)_L \times U(1)_Y$, the covariant derivative reduces to $D_\mu = \partial_\mu$ in the kinetic energy terms.  The Majorana mass matrix $M$ is a complex symmetric matrix with eigenvalues of ${\cal O({\cal M})}$ and violates the lepton number $L$ \footnote{The charge conjugate of a chiral fermion field appearing in the Majorana mass term is defined by ${\psi_R}^c \equiv C \overline{\psi_R}^T$.}. $Y_\nu$ is the $n \times n'$ matrix containing the  neutrino Yukawa couplings to the Higgs boson doublet $\widetilde{\phi} = i\tau_2 \phi^*$.

The effective Lagrangian which is valid at energies less than ${\cal M}$ 
is constructed by integrating out the heavy Majorana neutrino fields $N_R$. The effective Lagrangian has a power series expansion in 
$1/{\cal M}$ of the form 
\begin{equation}
\label{leff}
{\cal L}_{eff}= {\cal L}_{SM}+
     \frac{1}{{\cal M}}{\cal L}^{d=5} + \frac{1}{{\cal M}^2} {\cal L}^{d=6} +
     \cdots\  \equiv {\cal L}_{SM}+ {\cal \delta L}^{d=5} + {\cal \delta L}^{d=6} +
     \cdots\  ,
\end{equation}
where
${\cal L_{SM}}$ contains all
$SU(3)_C \times SU(2)_L \times U(1)_Y$ invariant operators of dimension $d\leq 4$. The gauge invariant operators
of dimension $d > 4$, constructed 
from the SM fields, account for the physics effects of the heavy Majorana neutrinos at energies   
$\le {\cal M}$.

The effective Lagrangian is defined through the effective action~\cite{Santa2},
\bea
\label{S0}
e^{i S_{eff}}=
\exp\left\{ i \int d^4 x \,{\cal L}_{eff}(x) \right\} \equiv
e^{i S_{SM}}\int {\cal D} N{\cal D} \overline{N}  e^{i S_N} ~,
\eea
obtained by functional integration over the heavy Majorana neutrino fields.

All contributions to the effective Lagrangian can be obtained by expanding 
the heavy neutrino propagator, contained in \eq{S0}, in a power series in $1/M$, 
\be
\label{expan}
\frac{1}{i\dvr -M}
= -\frac{1}{M}
- \frac{i\dvr}{M^2} +  \dots \, .
\ee

As shown in detail in Ref~\cite{BGJ}, the substitution of Eq.~\rfn{expan} into \eq{S0}, yields the terms of dimension $\le 6$, which suffice for the purposes of this work.

\subsection{ d=5 operator}
The first term in \eq{expan} yields the $d=5$ operator of the effective Lagrangian for the seesaw model,
\be
\label{L5}
{\cal {\delta L}}^{d=5} = 
\frac{1}{2}\,  c^{d=5}_{\alpha \beta}\,
\left(\overline{{\ell_L}^c_\alpha}\,  {\widetilde\phi}^*\right)  \, 
 \left( {\widetilde\phi}^\dag \, {\ell_L}_\beta \right) 
+ \mrm{h.c.}
\ee
where
\be
c^{d=5}_{\alpha \beta}=\left(Y_\nu^* 
\,\frac{1}{M} \,Y_\nu^\dagger \right)_{\alpha \beta}\,.
\ee

This is the well-known $\left(\Delta L=2\right)$ $d=5$ operator~\cite{Weinberg} that generates Majorana masses 
for the light weak doublet
neutrinos $\nu_L$ when the Higgs doublet develops a
non-zero vacuum expectation value $v/\sqrt{2}\simeq 174 \ \gev$,  
\be
\label{mass}
{m}_{\alpha\beta}= - \frac{v^2}{2}\left(Y_\nu^* 
\,\frac{1}{M} \,Y_\nu^\dagger \right)_{\alpha \beta}\, =
- \frac{v^2}{2} \left(c^{d=5}_{\alpha \beta}\right)\ .
\ee

\subsection{ d=6 operator}
The second term in \eq{expan} leads to the $d=6$ operator
\be
\label{L6}
{\cal {\delta L}}^{d=6} = i\,\left[ 
\overline{\ell_L} \,  \widetilde\phi\, Y_\nu \,   
\frac{\dvr}{M_i^2} \,\left(Y_\nu^\dagger\,{\widetilde\phi}^\dagger\,
\ell_L\right)\right]\, = c^{d=6}_{\alpha \beta} \left( \overline{\ell_L}_\alpha
\,  \widetilde\phi\right) i \dvr \left({\widetilde\phi}^\dagger\, {\ell_L}_\beta \right)\, ,
\ee
where
\be
c^{d=6}_{\alpha \beta} = \left( Y_\nu \frac{1}{M^2} 
Y_\nu^\dagger \right)_{\alpha \beta} \ .
\ee

While the $d=5$ operator of the effective Lagrangian
is the unique dimension-five operator
compatible with the gauge symmetries of the SM, there are many $d=6$ operators other than the one in \eq{L6}. The RG evolution of the operator
couplings from the putative high-energy scale where 
they are produced down to the electroweak scale will induce mixing
among all $d=6$ operators~\cite{more}. Nevertheless, unless very unnatural cancellations are present, our tree-level $d=6$ effective Lagrangian should be a tell-tale signature of the seesaw mechanism.

The effect of the $d=6$ operator in \eq{L6} is to renormalize the neutrino kinetic energy. By rotating to the basis
\be
\label{ren-wf}
\nu_\alpha^\prime =  
\left(\delta_{\alpha \beta}
+ \frac{v^2}{4} c^{d=6}_{\alpha
\beta} \right) \nu_\beta~\,,
\ee
where the kinetic energy is diagonal, the physical impact of the $d=6$ operator is transferred to the couplings of neutrinos to gauge bosons.
Since, in this work, the effective Lagrangian
is restricted to ${\cal O}(1/{\cal M}^2)$, the $d=6$ operator does not further modify the effects of the $d=5$ operator, 

Inclusion of the $d=6$ operator in the charged
current implies that the leptonic mixing matrix of the effective theory is given by
\be
U^{eff}_{\alpha i} = \left( \delta_{\alpha\beta} 
- \frac{v^2}{4}\,c^{d=6}_{\alpha \beta}\right) \,U_{\beta i}\,, 
\ee
where $U$ is the usual MNS lepton mixing matrix. Thus, neutrino oscillations are affected by the presence of the
$d=6$ operator. We note that the sensitivity of neutrino oscillations 
to more phases than just the ``CKM''-like phase, although with effects 
suppressed by powers of 
$1/{\cal M}^2$, has been pointed out already in Ref.~\cite{Endoh} in a general context. 

Phenomenological
bounds for the $d=6$ operator also can be found in the literature, as
this operator has been dealt with previously in the context of
theories with extra dimensions~\cite{Gouvea}.
For the particular case of a very short baseline $L\simeq 0$, the oscillation
probability depends on the coefficient $c^{d=6}_{\alpha \beta}$:
\be
P(\nu_\alpha\to\nu_\beta)= \left|\, \delta_{\alpha \beta} -\frac{v^2}{2}\,c^{d=6}_{\alpha\beta}
\,\right|^2\ .
\ee

From the results of the short baseline experiments~\cite{SBL}, we obtain a 
bound
on the seesaw scale,
\be
  Y_\nu/M \lesssim 10^{-4} \ {\gev}^{-1} \,,
\ee
which is many orders of magnitude weaker than that obtained from the 
d=5 operator, although independent from it.

\section{Parameter counting}

It is necessary and sufficient to consider our tree-level effective
Lagrangian,
\be
{\cal L}_{eff}^{d\le6}=  {\cal L_{SM}} - \frac{1}{4}
\left[ c^{d=5}_{\alpha \beta}\,\,
 \left( \overline{\widetilde{\ell_L}}_\alpha \, \vec{\tau} \, {{\ell_L}_\beta} \right) 
 \left( \widetilde{\phi}^\dagger \, \vec{\tau} \, {\phi}   \right)  + \mrm{h.c.}\,\right]
+i \, c^{d=6}_{\alpha \beta}\,\,
\left[ \overline{\ell_L}_\alpha \,  \widetilde\phi\,  
\dvr\left({\widetilde\phi}^\dagger\,
{\ell_L}_\beta \right)\right]\,,
\label{effL}
\ee
in order to take into account the leading low-energy 
effects related to leptogenesis. 

In this section we compare how many physical parameters are contained in the seesaw Lagrangian to those in the low-energy effective Lagrangian. The counting of the parameters is done by analyzing the symmetry structure of the theory~\cite{Santa1} and it is detailed in Ref.~\cite{BGJ}. 

First, let us analyze the high-energy seesaw Lagrangian in \eq{Lagrangian}
seesaw model. The number of physical parameters 
contained in the appears in  Table~1. For the general case of $n^\prime$ heavy neutrinos and $n$ light lepton doublets, there are $(n + n^\prime + n n^\prime)$ physical moduli and $n(n^\prime -1)$ physical phases in the Yukawa and Majorana mass matrices of the seesaw model.  
Of the real parameters, $n$ are the charged
lepton masses, $n$ are the light Majorana neutrino masses and $n^\prime$ are the heavy Majorana neutrino masses, whereas the remaining $(n n^\prime -n)$ 
are mixing angles. Table 1 also illustrates the counting for some specific number of generations.
\begin{table}[ht]
\begin{center}
Table 1: Seesaw Model\\
\end{center}
\begin{center}
\begin{tabular}{|cc|cc|}
\hline
\multicolumn{2}{|c|}{Generations}
&\multicolumn{2}{|c|}{${\cal L}_{SM}$ + ${\cal L}_{N_R}$}\\
\hline
\hline
$N_R$& $\ell_L$ & Moduli &Phases \\
\hline
$n'$& n &$n+n'+ n n'$ &$n(n'-1)$ \\
\hline
3 & 3 & 15 & 6  \\
\hline
2 & 3 & 11 & 3  \\
\hline
2 & 2 & 8 & 2  \\
\hline
\end{tabular}
\end{center}
\end{table}

Consider now the Lagrangian of the effective theory truncated at the $d=5$ operators for $n$ light active families. The number of physical parameters  is given on the Table 2. Note that, with an effective Lagrangian containing only the $d=5$ operator, information is lost: 
the number of physical low-energy parameters is 
not equal to the number of parameters of the high-energy seesaw
model for any value of $n^\prime$. 
For example, for $n=2$, the $d=5$ effective Lagrangian would contain only 
2 charged lepton masses, 2 neutrino masses and one mixing angle and one phase,  to be compared with the 8 moduli and 2 phases of the high-energy Lagrangian for $n=n^\prime =2$. 
\begin{table}[ht]
\begin{center}
Table 2: Effective Theory ($d\le 5$)
\end{center}
\begin{center}
\begin{tabular}{|c|cc|}
\hline
{Generations} & \multicolumn{2}{|c|}
{${\cal L}_{SM} + \delta{\cal L}^{d=5}$}\\
\hline
\hline
$\ell_L$ & Moduli &Phases \\
\hline
n  &$\frac{n(n+3)}{2}$ & $\frac{n(n-1)}{2}$ \\
\hline
3 & 9 & 3  \\
\hline
2 & 5 & 1  \\
\hline
\end{tabular}
\end{center}
\end{table}

As shown in Table 3, the addition of the $d=6$ operator allows to recover
the missing parameters, since the number of physical parameters in the low-energy effective Lagrangian $(d\le 6)$ equals the number of physical parameters in the seesaw Lagrangian of Table~1 if $n^\prime = n$.
When some extra symmetry or constraint is imposed in the high-energy Lagrangian (i.e.,  $n'<n$, degenerate heavy neutrinos,  etc.), 
the low-energy Lagrangian still has the same form, which appears 
paradoxical since now it contains a larger number of independent parameters 
than the high-energy theory.  The resolution of the paradox is 
that hypothetical low-energy measurements then are correlated.
\begin{table}[ht]
\begin{center}
Table 3: Effective Theory ($d\le 6$)
\end{center}
\begin{center}
\begin{tabular}{|c|cc|}
\hline
{Generations} & \multicolumn{2}{|c|}
{${\cal L}_{SM} + \delta{\cal L}^{d=5}+ \delta{\cal L}^{d=6}$}\\
\hline
\hline
$\ell_L$ & Moduli &Phases \\
\hline
n  & $ n(n+2)$ & $ n(n-1)$ \\
\hline
3 & 15 & 6 \\
\hline
2 & 8 & 2  \\
\hline
\end{tabular}
\end{center}
\end{table}

\section{Conclusions}
We have established a generic relationship between the seesaw model,  
including its leptogenesis-related parameters, 
and exotic low-energy neutrino couplings.
The physical consequences of the low-energy dimension $6$ operator 
are suppressed by two inverse powers of the large seesaw scale, 
and consequently, there is 
little practical hope to observe them, 
unless the seesaw scale turns out to be surprisingly small.

\end{document}